\documentclass[nohyperref]{article}

\usepackage{microtype}
\usepackage{graphicx}
\usepackage{subfigure}
\usepackage{booktabs} % for professional tables
\usepackage{xcolor}

\usepackage{hyperref}

\usepackage[accepted]{icml2023}

\usepackage{amsmath}
\usepackage{amssymb}
\usepackage{mathtools}
\usepackage{amsthm}

\usepackage[capitalize,noabbrev]{cleveref}

\theoremstyle{plain}

\theoremstyle{definition}

\theoremstyle{remark}

\usepackage[textsize=tiny]{todonotes}

\usepackage{booktabs}
\usepackage{graphicx,subfigure,epsfig,psfrag,xcolor}
\usepackage{amssymb}% http://ctan.org/pkg/amssymb
\usepackage{pifont}% http://ctan.org/pkg/pifont
\usepackage{multirow}
\newcommand{\cmark}{\ding{51}}%
\newcommand{\xmark}{\ding{55}}%

\icmltitlerunning{Machine Learning-guided Lipid Nanoparticle Design for mRNA Delivery}

\begin{document}

\twocolumn[
\icmltitle{Machine Learning-guided Lipid Nanoparticle Design for mRNA Delivery}



\begin{icmlauthorlist}
\icmlauthor{Daisy Yi Ding}{a}
\icmlauthor{Yuhui Zhang}{b}
\icmlauthor{Yuan Jia}{c}
\icmlauthor{Jiuzhi Sun}{c}
%\icmlauthor{Firstname5 Lastname5}{yyy}
%\icmlauthor{Firstname6 Lastname6}{sch,yyy,comp}
%\icmlauthor{Firstname7 Lastname7}{comp}
%\icmlauthor{Firstname8 Lastname8}{sch}
%\icmlauthor{Firstname8 Lastname8}{yyy,comp}
\end{icmlauthorlist}

\icmlaffiliation{a}{Department of Biomedical Data Science, Stanford University, Stanford, CA, USA}
\icmlaffiliation{b}{Department of Computer Science, Stanford University, Stanford, CA, USA}
\icmlaffiliation{c}{Department of Chemistry, Stanford University, Stanford, CA, USA}

\icmlcorrespondingauthor{Daisy Yi Ding, Yuhui Zhang, Yuan Jia, Jiuzhi Sun}{\{dingd, yuhuiz, yuanj2, js77\}@stanford.edu}
%\icmlcorrespondingauthor{Firstname2 Lastname2}{first2.last2@www.uk}

\icmlkeywords{Machine Learning, ICML}

\vskip 0.3in
]



\printAffiliationsAndNotice{}  % leave blank if no need to mention equal contribution

\begin{abstract}
While RNA technologies hold immense therapeutic potential in a range of applications from vaccination to gene editing, the broad implementation of these technologies is hindered by the challenge of delivering these agents effectively.
Lipid nanoparticles have emerged as one of the most widely used delivery agents, but their design optimization relies on laborious and costly experimental methods. 
We propose to in silico optimize LNP design with machine learning models.
On a curated dataset of 622 LNPs from published studies, we demonstrate the effectiveness of our model in predicting the transfection efficiency of unseen LNPs, with the multilayer perceptron achieving a classification accuracy of 98\% on the test set. 
Our work represents a pioneering effort in combining ML and LNP design, offering significant potential for improving screening efficiency by computationally prioritizing LNP candidates for experimental validation and accelerating the development of effective mRNA delivery systems.
\end{abstract}

\vspace{-3.3mm}
%Our work shows promise in the direction of ML-guided LNP design, where ML models are used to prioritize LNP candidates for experimental validation and improve screening efficiency.

%\vspace{-1mm}
\section{Introduction}
\label{introduction}

RNA-based technologies have the potential to transform life science research and medicine by enabling one’s own cells to transiently synthesize therapeutics, mechanistic probes and diagnostics. This science has created new opportunities for therapeutic vaccinations, protein replacement therapies, immunotherapy, gene editing and gene reprogramming \cite{kaczmarek2017advances,sahin2014mrna,chakraborty2017therapeutic}. However, the single most recognized obstacle to the broad implementation of RNA-based technologies is the delivery of the polar polyanionic RNA across non-polar biological barriers \cite{dowdy2017overcoming, stanton2018current}. Various agents have been developed to encapsulate RNA, among which lipid nanoparticles (LNPs) are one of the most widely investigated.

LNPs are often formulated with four components: (1) an ionizable or cationic lipid for complexing the polyanionic RNA, (2) a helper phospholipid to stabilize the LNP, (3) sterols for facilitating endosomal escape, and (4) lipid-anchored poly(ethylene glycol) (PEG) lipids for increasing circulation time \cite{eygeris2021chemistry}. The functionalities of LNPs are governed by the size, shape, charge, and ratio of each component which can be optimized for specific targeting and disease applications. Seminal studies have demonstrated that the chemical structures of LNP components can influence transfection efficacy and organ selectivity through structure-activity relationships (SAR) analysis. While the previous LNP development is highly fruitful, it often requires the combinatorial synthesis of individual lipids and pooled or individual evaluation of formulated LNPs, the process of which can be laborious and costly.

We hereby aim to expand the scope of traditional experimental methods for designing LNP delivery systems by leveraging the power of computational methods. We realized that LNP design is a suitable application for exploiting ML because it provides a set of modifiable characteristics. As experimental data in this field continues to grow, and ML techniques advance in modeling chemical structures, we see great potential in using ML to optimize the design of delivery agents. Specifically, we formulate the problem as a classification task where we use supervised ML models to predict the transfection efficiency of LNPs, i.e. the ability to successfully deliver mRNA molecules into target cells and facilitate their expression.

\iffalse
\begin{figure}[h]
\centering
\includegraphics[scale=0.5]{biology_LNP.pdf}
%\includegraphics[scale=0.69]{LNP_1.jpg}
%\includegraphics{LNP_2.jpg}
\caption{(A) Cellular uptake pathways of LNP-encapsulated mRNA. (B) LNPs are often formulated with the following four components: (1) helper phospholipid, (2) ionizable or cationic lipid, (3) cholesterol, and (4) lipid-anchored poly(ethylene glycol) (PEG). Figures taken from \cite{swingle2021lipid}.}
\label{fig:biology}
\end{figure}
\fi

\iffalse
\begin{figure}[h]
\centering
\includegraphics[scale=0.3]{LNP_1.jpg}
\caption{Cellular uptake pathways of LNP-encapsulated mRNA. Figure taken from \cite{swingle2021lipid}.}
\label{fig:uptake}
\end{figure}

\begin{figure}[h]
\centering
\includegraphics[scale=0.22]{LNP_2.jpg}
\caption{LNPs are often formulated with the following four components: (1) helper phospholipid, (2) ionizable or cationic lipid, (3) cholesterol, and (4) lipid-anchored poly(ethylene glycol) (PEG). Figure taken from \cite{swingle2021lipid}.}
\label{fig:lnp}
\end{figure}
\fi

We propose to use molecular representation learning techniques and classification models to predict the transfection efficiency of multicomponent LNPs based on their chemical structures.
We first curated a dataset of 622 LNPs along with their level of transfection efficiency from published studies \cite{liu2021membrane, zhang2020functionalized, li2015orthogonal, li2020combinatorial}.
%characterized by luciferase expression following treatment of the LNP-encapsulated mRNA \cite{liu2021membrane}.
When empirically evaluating different molecular representation learning methods, we found that the representations generated based on rules of chemical domain knowledge are sufficiently predictive, while those generated through large graph neural networks do not necessarily offer superior performance.
On the curated dataset, we showed that ML models effectively predict the transfection efficiency of unseen LNPs, with the multilayer perceptron achieving a classification accuracy of 98\% on the test set.
Such a model has the potential to significantly improve the screening efficiency by prioritizing LNP candidates for experimental validation.

To summarise, our main contributions in this paper include: (1) we propose to aid the design process of LNPs with ML by formulating it into a classification task where the input is the chemical structure of an LNP and the output is its transfection efficiency, and to our knowledge, our work represents the first attempt in this
direction; (2) we curate a dataset of LNPs along with their corresponding levels of transfection efficiency, and will make this dataset public to facilitate collaborative efforts towards advancing the state-of-the-art in the optimization of LNP delivery systems; (3) through proof-of-concept studies, we demonstrate the feasibility of in silico predicting the transfection efficiency of LNP for delivering mRNA, a direction that holds great promise for advancing the development of effective LNP-based mRNA therapeutics.

%\vspace{-1.6mm}
\section{Related Work}
\paragraph{Machine learning for molecular representation}
A variety of ML models have been proposed for molecular property prediction (such as absorption, distribution, metabolism, excretion, and toxicity of small molecules), ranging from support vector machine (SVM) \cite{zernov2003drug, alvarsson2016large, hou2007adme}, random forest (RF) \cite{zhang2007random, svetnik2003random}, to neural networks \cite{wu2018moleculenet, chen2018rise, rifaioglu2019recent, zhang2017machine}.  
Among them, graph neural network (GNN) has recently attracted considerable attention for its capability in learning representations directly from the chemical information encoded by molecular graphs \cite{wu2018moleculenet, sun2020graph, xiong2019pushing, rong2020self}.
Specifically, molecular graphs are natural representations of chemical structures: while a graph $G = (V, E)$ describes the connectivity relations between a set of nodes $V$ and a set of edges $E$, a molecule can be considered as a graph consisting of a set of atoms (nodes) and a set of bonds (edges).
GNN learns the representation of each atom in the molecule by aggregating the information from its neighboring atoms and connected bonds through message passing across the graph in a recursive process.
However, despite the attraction GNNs have gained in the community, it has also been shown that GNNs can be outperformed by traditional descriptor-based methods with SVM and RF in terms of prediction accuracy and computational efficiency, especially in the low-data regimes \cite{jiang2021could, mayr2018large}. 

%\vspace{-3.3mm}
\paragraph{Machine learning for the design of biomolecule delivery agents}
There has been previous work on leveraging ML to guide the design of another important type of delivery agents called adeno-associated virus (AAV) capsids \cite{ogden2019comprehensive}.
The paper characterizes single-codon substitutions, insertions, and deletions of AAV capsids across functions relevant to in vivo delivery of gene therapies.
It shows that the ML-guided design of AAV capsids outperforms random mutagenesis.
Though its goal is similar to ours as it aims to optimize the delivery agents for gene or mRNA therapies using ML, the data modality is very different: an AAV capsid contains three structural Cap proteins while an LNP often consists of lipids and cholesterol.
In addition, LNPs as delivery vehicles are superior to viral vectors such as AAVs in the sense that LNPs have lower immunogenicity and are capable of delivering larger payloads \cite{swingle2021lipid}.

%\vspace{-1.5mm}
\section{Approach}

\begin{figure*}[h!]
    \centering
    \includegraphics[width=0.85\linewidth]{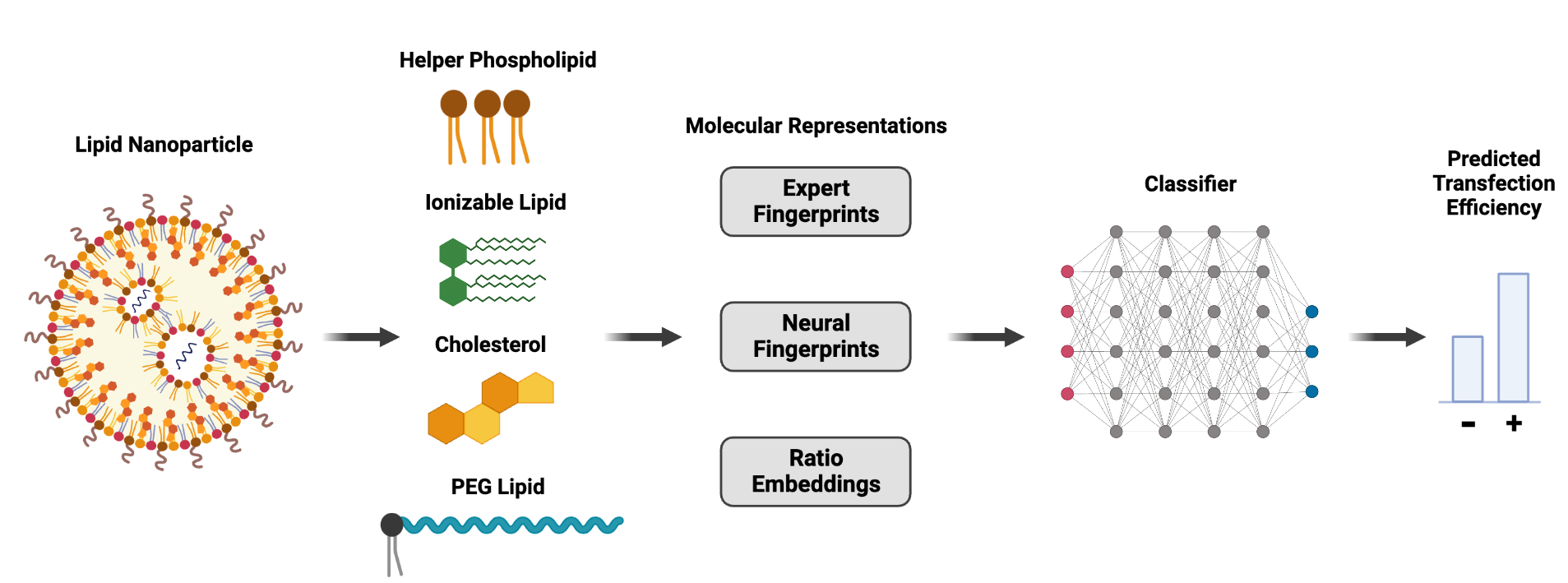}
    \vspace{-0.5em}
    \caption{Overview of our approach: (1) Molecular representation learning: we derive ``expert fingerprints" and ``neural  fingerprints" for each of the four molecular components of an LNP, and encode the mixture ratio with one-hot embeddings. (2) Downstream classification: we feed the concatenated molecular representations into  classifiers such as SVM or MLP to predict if the transfection efficiency of an LNP design is satisfying.
    This figure is generated with {\tt BioRender}.}   \label{fig:architecture}
    \vspace{-1em}
\end{figure*}

\subsection{Problem Formulation}
We formulate the LNP design task as a classification task. 
The input is a multicomponent LNP consisting of four molecules, along with the corresponding ratios in the LNP. 
The output is a label $y \in (0,1)$ indicating if the level of transfection efficiency is satisfying.

\vspace{-0.5mm}
\subsection{Model Design}
Our proposed model consists of two parts: (1) molecular representation learning and (2) downstream classification (Figure~\ref{fig:architecture}). 
Given an LNP with four components and their corresponding ratios, we first use molecular representation learning techniques to generate continuous low-dimensional representations to capture high-level information about the chemical structures and encode the ratio information with one-hot embeddings. 
We then feed the concatenated representations into a downstream classifier to predict the response, i.e. transfection efficiency.
The model performance is affected by both (1) the quality of the generated molecular representations in encoding information of the chemical structures; (2) the capability of the classifier in distinguishing satisfying and unsatisfying designs.
In this work, we explore molecular representations generated based on an expert-designed system and on GNNs, respectively, and compare classifiers including SVM, RF, XGBoost \cite{chen2016xgboost}, and multi-layer perceptron (MLP). 

%\vspace{-1mm}
\subsubsection{Molecular Representation Learning}

We propose to use two approaches to generate molecular representations to encode the structural and semantic information of chemical structures.
The first approach utilizes chemical knowledge by designing a set of rules through specified functions to extract different properties of the molecules such as atom and bond types.
The second approach treats each molecule as a heterogenous graph, where each atom is represented as a node and bond as an edge, and then leverages methods in graph representation learning to generate the representations through graph convolutions.
While the first approach using domain expertise is more interpretable, the second approach automates the learning of representations and mitigates the labor involved in handcrafting features.
Such representations are usually called molecular fingerprints in the field of chemoinformatics.
In this work, we denote the representations generated through the first approach as ``expert fingerprints", and through the second approach as ``neural fingerprints".

\vspace{-1mm}

\paragraph{Expert Fingerprint} 

% Introduce RDKit fingerprint here (chemistry knowledge)
Based on chemical knowledge, one can design a set of specified functions to extract important attributes of molecules that could affect their properties, such as atom types, bond types, chirality, functional groups, etc. 
These functions are carefully crafted by domain experts and have been made available in chemoinformatics softwares. 
%As these fingerprints are designed from chemical knowledge, we denote these fingerprints as ``expert fingerprints". 
We use RDKit\footnote{https://www.rdkit.org}, an open-source cheminformatics software, to generate fingerprints. 
RDKit provides a number of generation functions, and here we use RDKit2DNormalized, which is one of the most commonly used functions in previous works on molecular property prediction. 
The generated       ``expert fingerprints" have 200 dimensions.

\vspace{-1mm}
\paragraph{Neural Fingerprint}

Molecules can be viewed as heterogenous graphs consisting of different types of nodes (atoms) and edges (bonds). 
With graph representation learning, we can learn representations through graph convolutions in an automatic way. % without handcrafting fingerprints. 
%We name features extracted by graph neural networks as neural fingerprints.
Specifically, we use {\tt Grover}
%\footnote{https://papers.nips.cc/paper/2020/hash/94aef38441efa3380a3bed3faf1f9d5d-Abstract.html} 
to generate neural fingerprints as it has achieved state-of-the-art performance on various molecular prediction tasks \cite{rong2020self}. 
{\tt Grover} has its architecture based on one of the most expressive GNN variants {\tt GTransformer}. 
It is pre-trained on a large collection of molecules with self-supervised pre-training, which enables it to learn general representations. 
%The details of {\tt GTransformer} and the pre-training method are described in Appendix~\ref{sec:grover}.
%{\color{blue}{To do: add this from the course report to Appendix?}}. 
We compared {\tt Grover-base} and {\tt Grover-large}, which contain 48M and 100M parameters and output 3400- and 5000-dimension fingerprints, respectively\footnote{We only used {\tt Grover} to generate fingerprints and did not fine-tune its parameters due to the limited amount of training data.}.

\vspace{-1mm}

\paragraph{Component Ratio Embedding}

The ratio of each molecular component in an LNP can significantly affect its functionality \cite{kauffman2015optimization, hassett2019optimization}. 
To encode the ratio information, we normalize the ratio of each component to $(0, 1)$ and round it to the nearest number in $\{0.05, 0.10, ..., 1.00\}$.
We then use one-hot embeddings to encode the ratio information of each component, each of which has 20 dimensions. 

\subsubsection{Classification Models}
With the generated vector representations of a multicomponent LNP, we concatenate the representations of the different components and their ratios and feed the aggregated representation into a downstream classifier of choice.
We experiment with classification models including SVM, RF, XGBoost, and MLP.
The classifier learns to differentiate the representations through supervised learning and outputs the likelihood of a molecule being capable of achieving satisfying transfection efficiency. 

\subsubsection{Evaluation Metric}
We use the area under the receiver operating characteristic (AUROC) as our evaluation metric, which measures how well a model ranks examples and distinguishes between classes (i.e. satisfying versus unsatisfying designs of LNPs).
%Since our final goal is only to differenciate the satisfied LNP designs with non-satisfied ones instead of caring whether the likelihood estimation from the classifier is accurate, we use AUC-ROC as our evaluation metric (TODO: Daisy). 

%\vspace{-0.5mm}

\section{Experiments and Discussion}
\vspace{-0.5mm}
\subsection{Data Curation}
%\vspace{-1mm}
We curated a dataset of 622 LNPs along with their level of transfection efficiency based on published studies \cite{liu2021membrane, zhang2020functionalized, li2015orthogonal, li2020combinatorial}, as there were no publicly available data resources that can be readily used to develop ML models for LNP design.
%on membrane-destabilizing ionizable phospholipids for mRNA delivery 
%The paper characterizes the functional capability of a set of multi-tailed ionizable phospholipids (iPhos) generated through combinatorial synthesis.
%iPhos lipids function with other helper lipids in formulating the multicomponent LNPs for mRNA delivery, with structure-activity relationships demonstrating that the chemical structure of iPhos can control in vivo efficacy.
%We decided to curate this dataset from scratch because we did not find any publicly available resources that can be readily used to develop ML models for LNP design.
%We chose the data from \cite{liu2021membrane} because it provides us with a relatively large-scale dataset by itself.
%We have also considered combining multiple datasets (e.g. from multiple papers) to increase the data scale, but we want to note the challenge of normalizing datasets from different sources.
%For example, different research labs may use different approaches to measure transfection efficiency or use different cell lines in their experiments, which make datasets heterogeneous and confounded by other factors.
We represent the chemical structures of the four components in an LNP with SMILES (Simplified Molecular Input Line Entry System) codes, which provide 2D representations of the atoms and the bonds in molecules.
To generate the SMILES codes, the chemical structures were first drawn by domain chemists in a software called {\tt ChemDraw}.

Due to variations in transfection efficiency readouts across different studies \cite{liu2021membrane, zhang2020functionalized, li2015orthogonal, li2020combinatorial}, domain scientists have categorized LNP designs into two groups: those with satisfying transfection efficiency (141 LNPs) and those with unsatisfying transfection efficiency (481 LNPs).
This categorization was done to facilitate model development on an integrated dataset. 
However, it is important to acknowledge the potential bias introduced in this categorization process, given that the readouts cannot be directly compared across different studies.

%with satisfying and unsatisfying transfection efficiency (comprising 141 and 481 LNPs, respectively) to enable model development on the integrated dataset.
%However, we also want to note the potential bias in categorizing LNPs since the readouts are not directly comparable across papers.
%We then obtained the readouts on transfection efficiency from the results presented in \cite{liu2021membrane}, where the readouts are characterized with luciferase expression following treatment of IGROV1 cells with iPhos-delivered mRNA.
%We used a threshold of 10,000 relative light units to distinguish between LNP designs with satisfying and unsatisfying transfection efficiency, comprising 91 and 481 LNPs, respectively.

The availability of large-scale, high-quality data is often a bottleneck for biomedical applications, including for the optimization of LNP delivery systems. 
In this study, we address the challenge by curating this new dataset, and by making this dataset publicly available, we aim to contribute to the scientific community and enable more efficient LNP optimization with ML techniques. 
%The integration of curated datasets with ML approaches holds great promise for advancing the field of LNP design and facilitating the development of effective therapeutic interventions.
%As large-scale high-quality data are often the key bottleneck for biomedical applications, we hope that our efforts in curating this new dataset can contribute to the community by enabling more efficient LNP optimization with ML.
%\footnote{We want to acknowledge that Yuan Jia and Gillian Sun helped us tremendously in the data curation process. This project would not be possible without their help and expertise in chemistry.}.

%\subsection{Data Split}
\vspace{-1.5mm}
\subsection{Experiment Setup}
%\vspace{-1mm}

% We split the dataset into training, validation, test, and external test sets, with 497, 62, 63, and 65 data points in each set, respectively (roughly corresponding to a ratio of 7/1/1/1). {\color{blue}{To do: change here as we are removing the external test set}}. 
We split the dataset into training, validation, and test sets, with 497, 62, 63 data points in each set, respectively (roughly corresponding to a ratio of 8/1/1). 
For SVM, we experiment with different kernels, including the linear kernel, the radial basis function kernel, and the polynomial kernel, and different regularization strengths.
For RF, we experiment with different numbers of trees in the forest and different values for the max depth of each tree.
For XGBoost, we experiment with different number of estimators and  regularization strengths.
For MLP, we use one hidden layer with 128 neurons and train the model with the Adam optimizer.
We use the validation set to select the optimal hyperparameters for each model. 

\vspace{-1.5mm}
\subsection{Results}
%\vspace{-1mm}

Experimental results of different classifiers trained with different molecular fingerprints (FP) are shown in Table~\ref{tab:results}. 
Our methods achieve overall satisfying performances on the LNP transfection efficiency prediction task, with the MLP trained with ``expert fingerprints" achieving the highest predictive accuracy of 98\% on the test set.
Specifically, ``expert fingerprints" generated based on domain knowledge are sufficiently predictive by capturing the structural and semantic information of the molecules, while ``neural fingerprints" generated with GNN ({\tt Grover}) do not necessarily offer superior performance. 
%Expert fingerprints generated with a set of specified functions based on chemical knowledge are generally good enough to capture the structual and semantic information of the molecules. 
%Meanwhile, neural fingerprints generated with GNN ({\tt Grover}) do not show significantly better performance compared to the expert fingerprints.
%The same findings applied to the results from the PCA analysis, we find that the information is well captured by expert finerprints, and neural fingerprints generated by the Grover base and Grover large does not seem to provide additional information. 
This finding is consistent with those presented in \citet{jiang2021could}. 
It may also be caused by the fact that the molecules in LNPs are generally larger than those used to train {\tt Grover}. 
Since ML often does not perform well on out-of-distribution data, such distribution differences in the size of the molecules could be the reason for GNNs not generating better molecular representations. 
Overall, our proof-of-concept studies demonstrate the feasibility of using ML to model the chemical structure of multicomponent LNPs in predicting their functionality and show promise in ML-guided LNP design.

\begin{table}[t]
    \small
    \centering
    \begin{tabular}{l|ll|ll}
        \toprule
        Classifier & Expert FP & Neural FP & Val AUC & Test AUC \\
        \midrule
        \multirow{4}{*}{SVM} & \cmark & \xmark & 0.9092 & 0.9614 \\
         & \xmark & \cmark & 0.9333 & 0.9627 \\
         &  \cmark & \cmark & \textbf{0.9518} & \textbf{0.9654} \\
        & \cmark & \cmark (Large) & 0.9219 & 0.9481 \\
        \midrule
         & \cmark & \xmark & 0.9326 & \textbf{0.9521} \\
        Random & \xmark & \cmark & \textbf{0.9361} & 0.9208 \\
        Forest & \cmark & \cmark & 0.9255 & 0.9215 \\
        & \cmark & \cmark (Large) & 0.9355 & 0.9375 \\
        \midrule
        \multirow{4}{*}{XGBoost} & \cmark & \xmark & \textbf{0.9660} & \textbf{0.9468} \\
         & \xmark & \cmark & 0.9489 & 0.9348 \\
         & \cmark & \cmark & 0.9560 & 0.9402 \\
         & \cmark & \cmark (Large) & 0.9390 & 0.9282 \\
        \midrule
        \multirow{4}{*}{MLP} & \cmark & \xmark & \textbf{0.9617} & \textbf{0.9815} \\
         & \xmark & \cmark & 0.9300 & 0.8892 \\
         & \cmark & \cmark & 0.9383 & 0.9169 \\
         & \cmark & \cmark (Large) & 0.9533 & 0.9508 \\
        \bottomrule
    \end{tabular}
     \vspace{-1em}
    \caption{Experiment results. Transfection efficiency can be accurately predicted by ML approaches using ``expert fingerprints" on our curated dataset.}
    \label{tab:results}
     \vspace{-1.8em}
\end{table}

% \begin{table}[htbp]
%     \small
%     \centering
%     \begin{tabular}{c|cc|cc|cc}
%         \toprule
%         Method & Val F1 & Val AUC & Test F1 & Test AUC & Ext F1 & Ext AUC \\
%         \midrule
%         \multicolumn{7}{c}{497/62/63/65 For Train/Val/Test/Ext} \\ 
%         MLP+RDKitFP(200*4+20*4) & 0.7619 & 0.9617 & 0.5556 & 0.9815 & 0.0000 & 0.8712 \\
%         MLP+GroverBaseFP(3400*4+20*4) & 0.7143 & 0.9300 & 0.6875 & 0.8892 & 0.7742 & 0.8511  \\
%         MLP+RDKitFP+GroverBaseFP(200*4+3400*4+20*4) & 0.7000 &  0.9383 & 0.4706 & 0.9169 & 0.0000 & 0.8499\\
%         MLP+RDKitFP+GroverLargeFP(200*4+5000*4+20*4) & 0.5882 & 0.9533 & 0.2667 & 0.9508 & 0.0000 & 0.7506\\
%         % \multicolumn{3}{c}{115 Data} \\
%         % MLP+dFP & 0.7262 & 0.7391 \\
%         % MLP+cFP & 0.7386 & 0.7391 \\
%         \bottomrule
%     \end{tabular}
%     \caption{TODO: remove F1. We only care about if the model can differenciate high score with low score, we dont need to force the threshold to be 0.5.}
%     \label{tab:my_label}
% \end{table}

\vspace{-1.5mm}
\section{Conclusion}
%\vspace{-1mm}

In this work, we proposed to use ML models to predict the transfection efficiency of LNPs for mRNA delivery.
We curated a dataset from scratch and empirically evaluated various molecular representation learning techniques and classification models on the task. 
Our methods achieved overall satisfying performance, demonstrating the feasibility of leveraging ML to model the chemical structures of LNPs in predicting their functionality. 
Our work shows promise in the direction of ML-guided LNP design, where ML models are used to prioritize LNP candidates for experimental validation and improve screening efficiency.
As for future directions, we will leverage the developed ML model to predict and prioritize new LNP designs based on their probability of achieving high transfection efficiency and conduct wet-lab experiments to further validate the effectiveness of our model. 
%With further data input and training, 
We aspire for this prototype to evolve into a powerful tool for assisting researchers in the systematic in silico selection of promising LNP designs.
%{\color{blue}{To do: update}}
%Moreover, while our external test set is curated from the same source as the training data in this work, we plan to examine the generalizability and robustness of our model more rigorously by testing it on the datasets curated from other sources.

\section*{Reproducibility Statement}

We provide all the code and data in \url{https://anonymous.4open.science/r/Lipid_Nanoparticle_Design/}. 
This will help researchers reproduce the results while enabling further exploration of other methods and directions on the curated dataset.
%extend other directions with our provided data.

\small{
\bibliography{example_paper}
\bibliographystyle{icml2023}
}

\end{document}